\documentclass[12pt]{iopart}
\usepackage{iopams}
\usepackage{epsf}
\begin{document}

\title[$\rho(770)^0$ and $f_{0}(980)$ Production in Au+Au and pp
Collisions at $\sqrt{s_{NN}}$ $\!=\!$ 200 GeV]{$\rho(770)^0$ and
$f_{0}(980)$ Production in Au+Au and pp Collisions at
$\sqrt{s_{NN}}$ $\!=\!$ 200 GeV}

\author{Patricia Fachini\dag\ for the STAR Collaboration\ddag
\footnote[3]{To whom correspondence should be addressed
(pfachini@bnl.gov)} }

\address{\dag\ Brookhaven National Laboratory, Bldg. 510A, Upton, NY 11973-5000,
USA.}

\address{\ddag\ See Ref. \cite{1} for full collaboration list.}

\begin{abstract}
Preliminary results on $\rho(770)^0$ and $f_{0}(980)$ production
at mid-rapidity using the mixed-event technique are presented. The
$\rho^0$ and $f_0$ measurements via their hadronic decay channel
in minimum bias pp and peripheral Au+Au collisions at
$\sqrt{s_{NN}}$ $\!=\!$ 200 GeV were performed using the STAR
detector at RHIC. This is the first direct measurement of
$\rho^0(770) \!\rightarrow\! \pi^+\pi^-$ and $f_0(980)
\!\rightarrow\! \pi^+\pi^-$ in heavy-ion collisions. The $\rho^0$
mass is transverse momentum dependent and significantly shifted in
minimum bias pp and peripheral Au+Au interactions. The
modifications of masses, widths and line shapes of short-lived
resonances due to phase space and dynamical effects are discussed.
The $\rho^0/\pi^-$ and the $f_0/\pi^-$ ratios are compared to
measurements in pp and e$^{+}$e$^{-}$ interactions at various
energies.
\end{abstract}

% Uncomment for PACS numbers title message


\maketitle

\section{Introduction}
The measurement of resonances with lifetimes smaller than, or
comparable to, the lifetime of the dense matter produced in
relativistic heavy-ion collisions provides an important tool for
studying the collision dynamics \cite{2}. Physical effects such as
thermal weighting of the states \cite{3,4,5,6,7} and dynamical
interactions with matter \cite{5,7} may modify resonance masses,
widths and shapes.

Partial-waves analysis have successfully parameterized $\pi \pi$
scattering \cite{8}. Introducing to the formalism the rescattering
of pions, in which $\pi^{+}\pi^{-} \!\rightarrow\! \rho(770)^0$,
the interference between different scattering channels can distort
the line shape of resonances \cite{9}.

Due to the proximity to the K $\!\bar{\textrm{K}}$ threshold, the
$f_0(980)$ resonance has been interpreted as a K
$\!\bar{\textrm{K}}$ molecule (bound state) \cite{10}. The study
of this exotic state might give insight on the strangeness
behavior at the various stages of the collision.

Previous measurements of the $\rho^0$ meson in hadronic $Z^0$
decays indicate that the $\rho^0$ line shape is considerably
distorted from a relativistic p-wave Breit-Wigner shape,
especially at relatively low momentum in multipion systems
\cite{11,12,13}. A mass shift that may be -30 MeV/$c^{2}$ or
larger was observed.

In pp collisions, the $\rho^0$ meson has been measured at
$\sqrt{s}$ $\!=\!$ 27.5 GeV \cite{14}. This is the only pp
measurement used in the hadroproduced $\rho^0$ mass average
reported in \cite{15}. In this measurement, the $\rho^0$ mass
$m_{\rho}$ $\!=\!$ 0.7626 $\!\pm\!$ 0.0026 GeV/$c^2$ was obtained
from a fit to a relativistic p-wave Breit-Wigner function times
the phase space \cite{14}.

Preliminary results from the first measurement of $\rho^0(770)$
and $f_{0}(980)$ via their hadronic decay channel in minimum bias
pp and peripheral Au+Au collisions at $\sqrt{s_{NN}}$$\!=\!$ 200
GeV using the STAR (Solenoidal Tracker At RHIC) detector at RHIC
are presented.

\section{Data Analysis}
The main STAR detector consists of a large Time Projection Chamber
(TPC) \cite{16} placed inside a uniform solenoidal magnetic field
of 0.5 T that provides the measurement of charged particles. About
11 million minimum bias pp events and about 1.2 million events for
the 40$\%$ to 80$\%$ of the hadronic Au+Au cross section at
$\sqrt{s_{NN}}$ $\!=\!$ 200 GeV were used in this analysis. Since
the pion daughters from $\rho^0$ decays originate at the
interaction point, only tracks whose distance of closest approach
to the primary interaction vertex was less than 3 cm were
selected. Charged pions were selected by requiring their energy
loss ($dE/dx$) in the gas of the TPC to be within three standard
deviations (3$\sigma$) of the expected mean. In order to enhance
track quality \cite{17}, the daughters were also required to have
pseudorapidities $|\eta| \!<\! 0.8$ and transverse momenta ($p_T$)
greater than 0.2 GeV/$c$.

The $\rho^0$ and $f_0$ measurements were performed calculating the
invariant mass for each oppositely charge $\pi \pi$ pair in an
event. The resulting invariant mass distribution was then compared
to a reference distribution calculated from the geometric mean of
the invariant mass distributions obtained from $\pi^+ \pi^+$ and
$\pi^- \pi^-$ pairs from the same event.

\begin{figure}[htb]
\begin{minipage}[t]{80mm}
\begin{center}
\epsfxsize=3.0in
\epsfbox{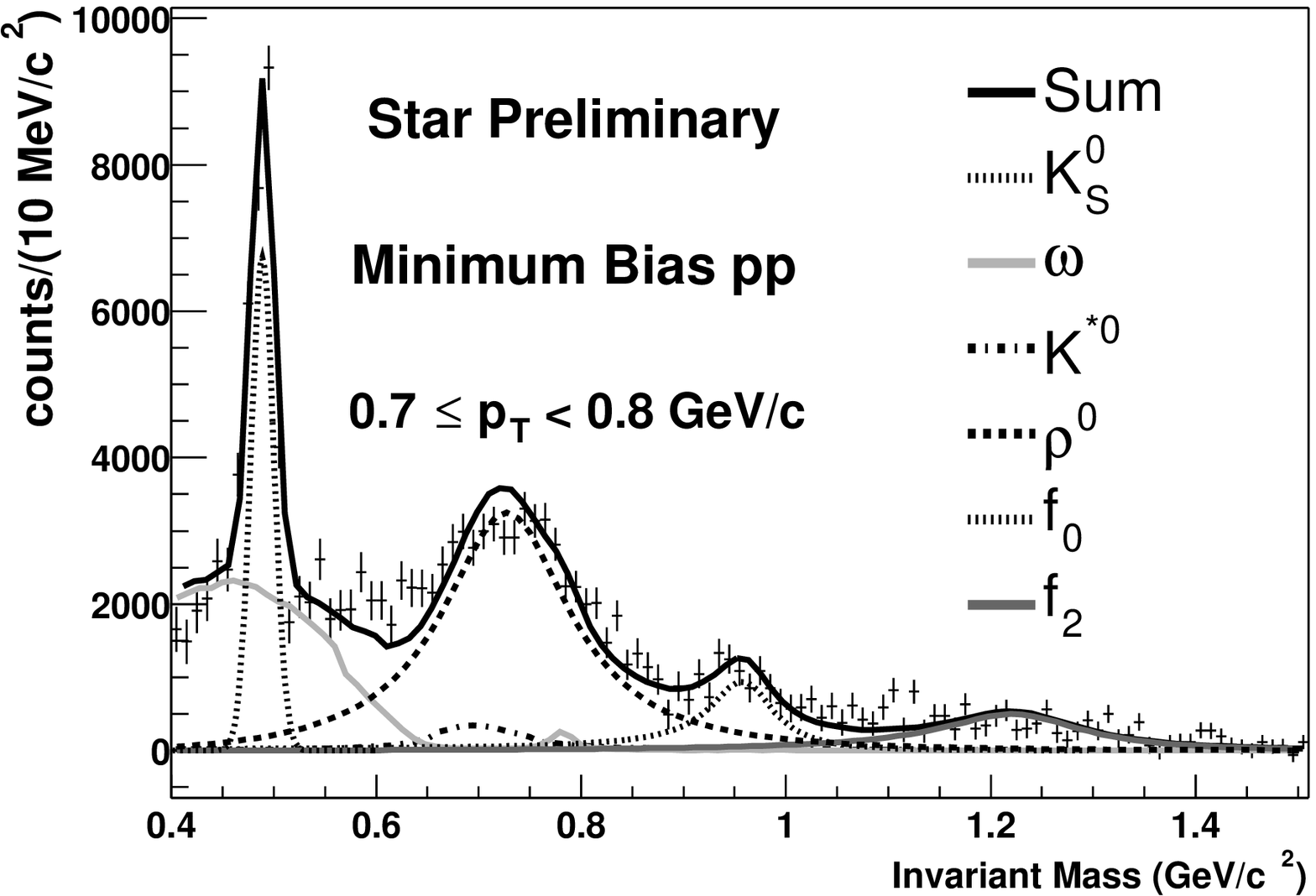}
\end{center}
\end{minipage}
\hspace{\fill}
\begin{minipage}[t]{80mm}
\begin{center}
\epsfxsize=3.0in
\epsfbox{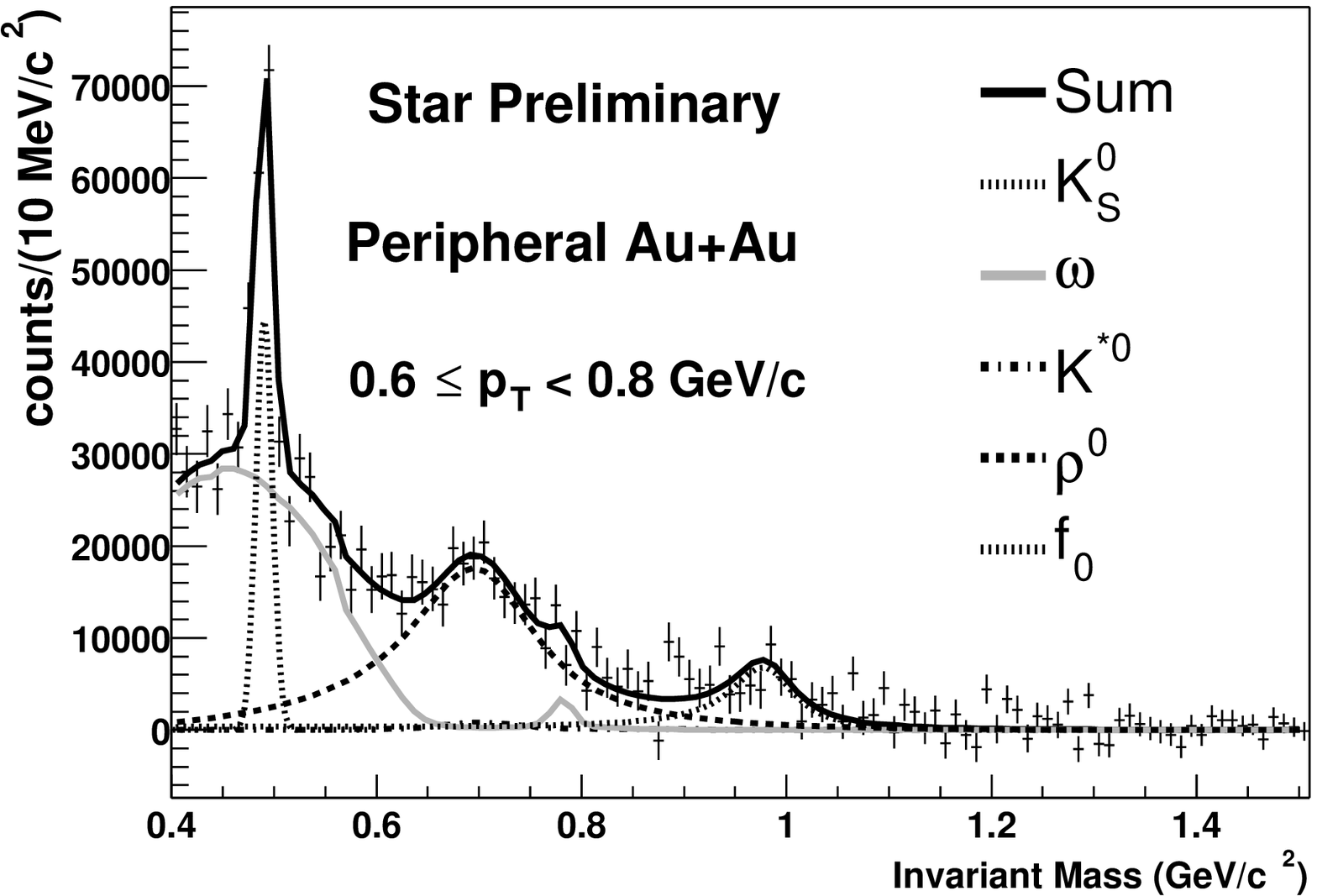}
\end{center}
\end{minipage}
\caption{\label{fig:cocktail}The raw $\pi^+\pi^-$ invariant mass
distributions after subtraction of the like-sign reference
distribution for minimum bias pp (left) and peripheral Au+Au
(right) interactions. In minimum bias pp, the $\pi^+ \pi^-$
Bose-Einstein correlation is accounted for and affects only
$M_{\pi\pi} \!<\!$ 0.45 GeV/$c^2$. For this particular $p_T$ bin,
the detector efficiency and acceptance correction for the
invariant mass region shown is approximately constant and
$\sim$25$\%$ for minimum bias pp and varies from $\sim$25$\%$ to
$\sim$35$\%$ for peripheral Au+Au collisions.}
\end{figure}

The $\pi^+\pi^-$ invariant mass distribution ($M_{\pi\pi}$) and
the like-sign reference distribution were normalized to each other
at $M_{\pi\pi} \!\gtrsim\!$ 1.5 GeV/$c^2$. The resulting raw
distributions after subtraction for minimum bias pp and peripheral
Au+Au collisions at mid-rapidity ($|y| \!\leq\! 0.5$) for a
particular $p_T$ bin are shown in Fig.~\ref{fig:cocktail}. The
$p_T$ coverage of the $\pi^+\pi^-$ pair is 0.2 $\!\leq\! p_T
\!\leq\!$ 2.2 GeV/$c$ for minimum bias pp and peripheral Au+Au
collisions. The hadronic ``cocktail" was described as follows. The
$K_S^0$ was fit to a gaussian (dotted line). The $\omega$ (light
grey line) and $K^{\ast}(892)^{0}$ (dash-dotted line) functions
were obtained from the HIJING event generator \cite{18}, with the
kaon being misidentified as a pion in the case of the $K^{\ast
0}$. The $\rho^0(770)$ (dashed line), the $f_0(980)$ (dotted line)
and the $f_2(1270)$ (dark grey line) were fit to the relativistic
Breit-Wigner function (BW) \cite{19} times the Boltzman factor
\cite{4,5,6,7} PS $\!=\!$ $(M_{\pi\pi}/\sqrt{M_{\pi\pi}^2 \!+\!
p_T^2})\cdot\exp(\!-\!\sqrt{M_{\pi\pi}^2 \!+\! p_T^2}/T)$ that
represents the phase space. Here, $T$ is the temperature in which
the resonance is emitted \cite{5}. The uncorrected numbers of
$\rho$, $f_0$, $\omega$, $K_S^0$ and $f_2$ were free parameters in
the fit while the $K^{\ast 0}$ fraction was fixed according to the
$K^{\ast}(892)^{0} \!\rightarrow\! \pi$K measurement \cite{20}.
The $\rho^0$, $\omega$ and $K^{\ast 0}$ functions were corrected
for the detector acceptance \cite{17}. The solid black line in
Fig.~\ref{fig:cocktail} is the sum of all the contributions in the
hadronic ``cocktail".

\begin{figure}[htb]
\begin{minipage}[t]{80mm}
\begin{center}
\epsfxsize=2.9in \epsfbox{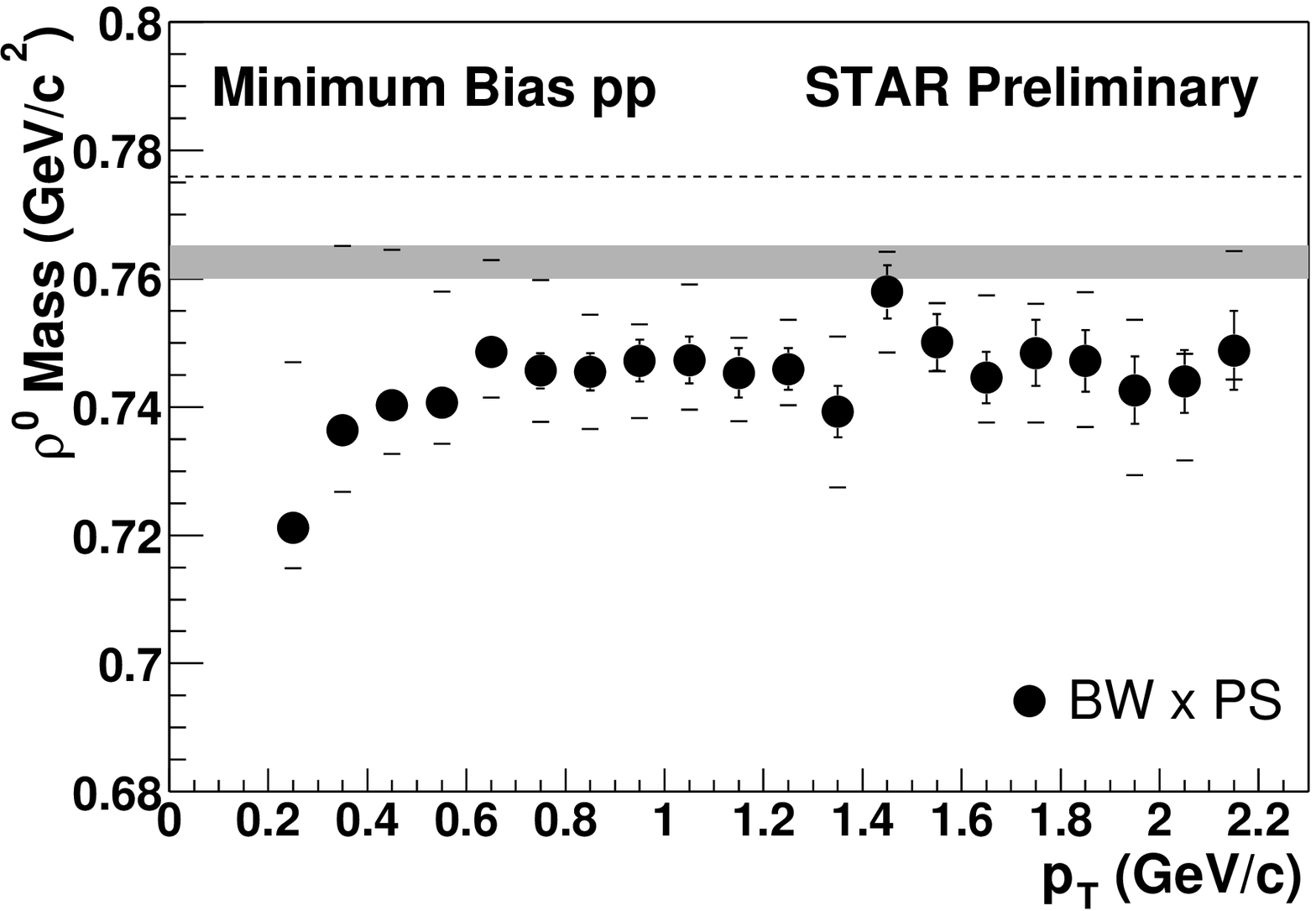}
\end{center}
\end{minipage}
\hspace{\fill}
\begin{minipage}[t]{80mm}
\begin{center}
\epsfxsize=2.9in \epsfbox{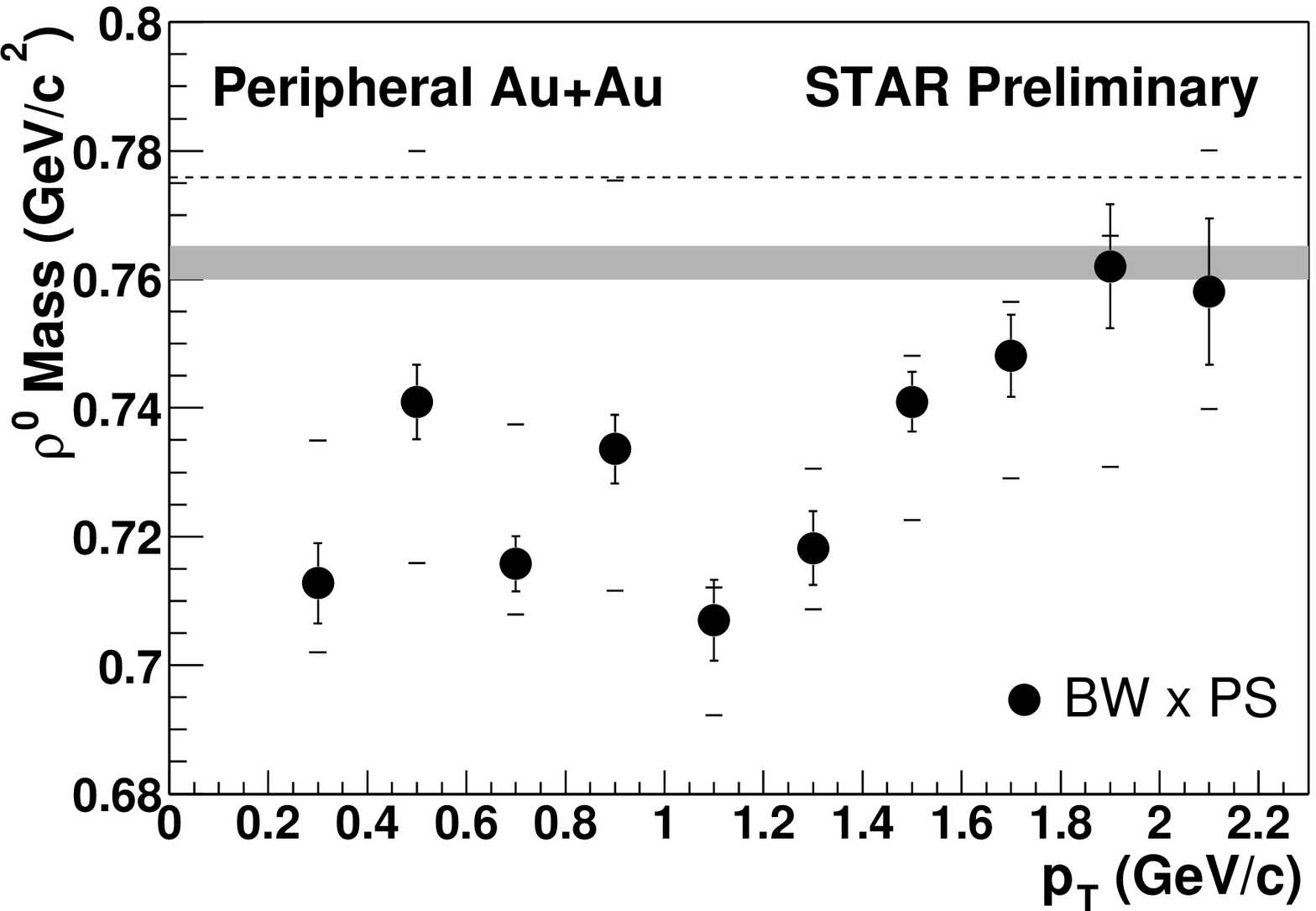}
\end{center}
\end{minipage}
\caption{\label{fig:mass}The $\rho^0$ mass as a function of $p_T$
for minimum bias pp (left) and for peripheral Au+Au (right)
collisions. The $\rho^0$ mass was obtained by fitting the data to
the BW$\times$PS function in the hadronic ``cocktail". The
horizontal solid lines represent the systematic uncertainties. The
shaded areas depict the $\rho^0$ mass measured in pp collisions
\cite{14}. The dashed lines sketch the average of the $\rho^0$
mass measured in e$^+$e$^-$ \cite{15}.}
\end{figure}

The $\rho^0$ mass as a function of $p_T$ for peripheral Au+Au and
minimum bias pp interactions are presented in Fig.~\ref{fig:mass}.
The $\rho^0$ mass was obtained by fitting the data to a
relativistic p-wave ($\ell$ $\!=\!$ 1) Breit-Wigner function times
the phase space (BW$\times$PS), represented by the filled circles
in Fig.~\ref{fig:mass}. The $\rho^0$ peak was also fit only to a
relativistic p-wave Breit-Wigner function; however, the fit lacked
in reproducing the $\rho^0$ line shape and usually underestimated
the position of the peak mainly at low $p_T$. The $\rho^0$ meson
width was fixed at $\Gamma_0$ $\!=\!$ 160 MeV$/c^2$, which is
consistent with the $\rho^0$ natural width \cite{15} folded with
the detector resolution. In this measurement, we do not have
sensitivity for a systematic study of the $\rho^0$ width. In Au+Au
collisions, the temperature used in the PS factor was $T$ $\!=\!$
120 MeV \cite{5}, while in pp $T$ $\!=\!$ 160 MeV \cite{21}. The
$\rho^0$ mass at mid-rapidity ($|y| \!\leq\!$ 0.5) for minimum
bias pp and peripheral Au+Au collisions at $\sqrt{s}$ $\!=\!$ 200
GeV increases as a function of $p_T$ and is systematically lower
than the value reported by \cite{15}. In addition, the $\rho^0$
mass measured in peripheral Au+Au collisions is lower than the pp
measurement. In this measurement, we do not have sensitivity for a
systematic study of the $f_0$ mass and width. In both minimum bias
pp and peripheral Au+Au collisions, the $f_0$ mass and width were
fixed at $\sim$980 MeV/$c^2$ and 75 MeV/$c^2$, respectively.

The hadronic ``cocktail" presented in Fig.~\ref{fig:cocktail}
depicts our best understanding of the data. However, the
correlations near the $\rho^0$ mass may be of unknown origins,
e.g. not well known particles, such as the $f_0(600)$ \cite{15}.
The uncorrected number of $\omega$ as a free parameter in the
hadronic ``cocktail" may be accounting for some unknown
contributions and causing the apparent increase in the
$\rho^0/\omega$ ratio between peripheral Au+Au and minimum bias pp
interactions. In order to evaluate the systematic uncertainty in
the $\rho^0$ mass due to unknown contributions in the hadronic
``cocktail", the $\rho^0$ mass was obtained by fitting the peak to
the BW$\times$PS function plus an exponential function
representing the unknown contributions. This uncertainty is
correlated between minimum bias pp and peripheral Au+Au and
corresponds to the main contribution to the systematic
uncertainties shown in Fig.~\ref{fig:mass}. The other contribution
of $\sim$3 MeV/$c^2$ corresponds to the uncertainty in measuring
the particle momentum. The mass resolution in the TPC is $\sim$9
MeV/$c^2$. The $\rho^0$ mass obtained from the fit using the
BW$\times$PS function plus an exponential function was always
higher than the mass obtained from the BW$\times$PS function in
the hadronic ``cocktail" fit.

\begin{figure}[htb]
\begin{minipage}[t]{80mm}
\begin{center}
\epsfxsize=2.8in \epsfbox{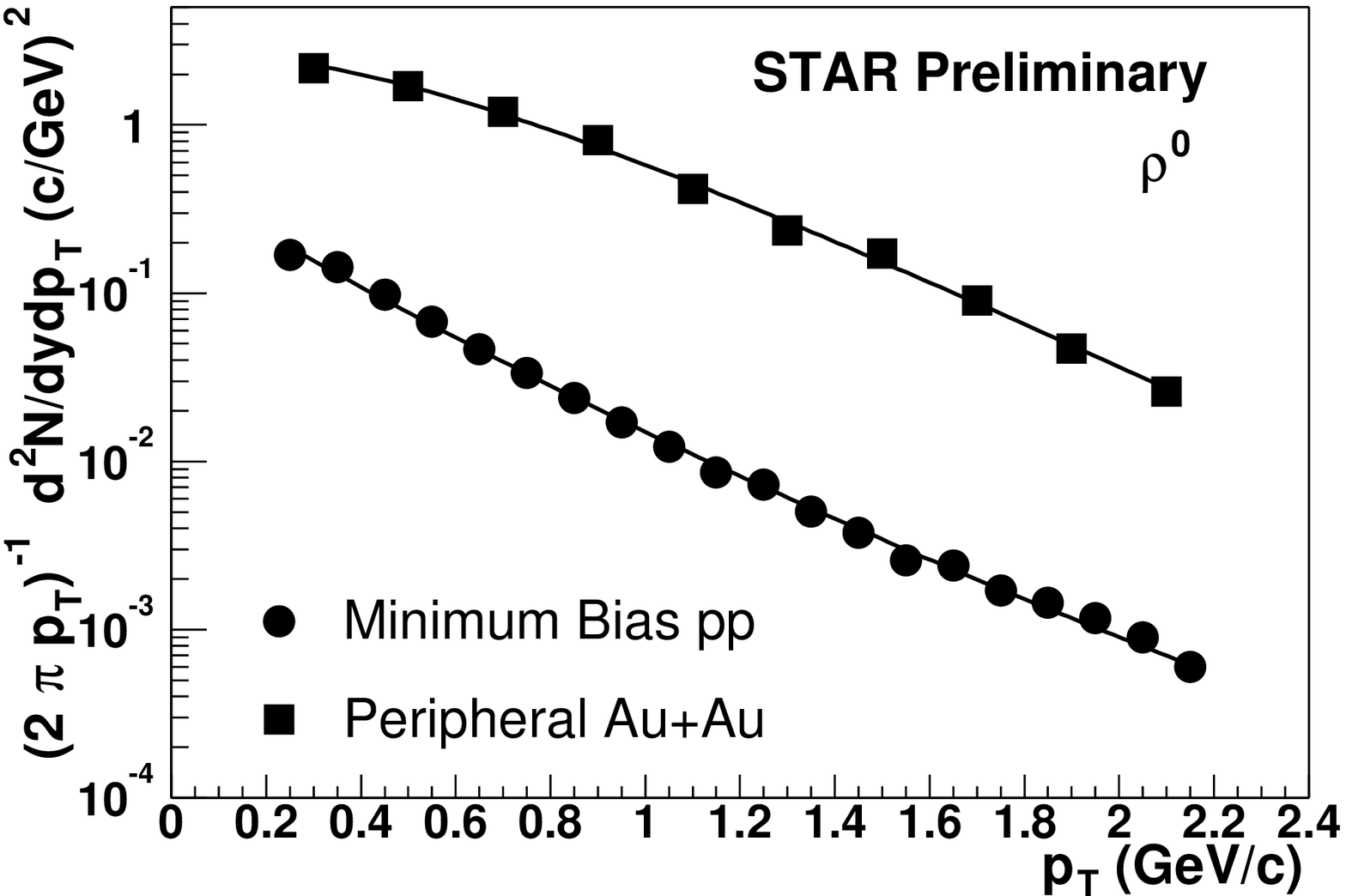}
\end{center}
\end{minipage}
\hspace{\fill}
\begin{minipage}[t]{80mm}
\begin{center}
\epsfxsize=2.8in \epsfbox{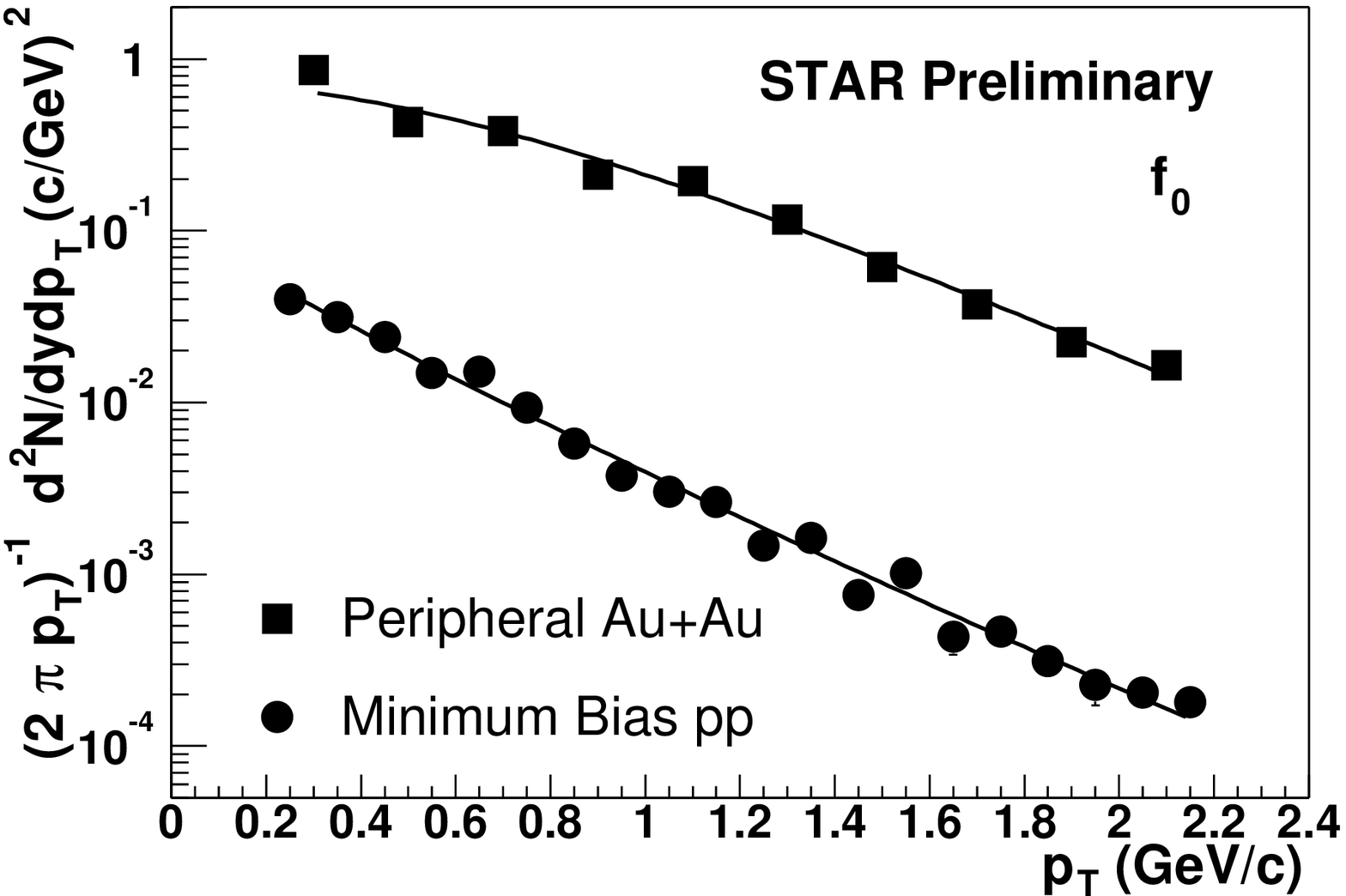}
\end{center}
\end{minipage}
\caption{\label{fig:spectra}The $p_T$ distributions at
mid-rapidity ($|y| \!<\!$ 0.5) from $\rho^0$ (left) and $f_0$
(right) produced in minimum bias pp and peripheral Au+Au
collisions. See text for explanation on the functions used to fit
the data. The errors shown are statistical only, and they are
smaller than the symbols that represent the measurements.}
\end{figure}

In order to obtain the resonance yield, detector acceptance and
efficiency corrections \cite{17} were applied to the uncorrected
numbers of $\rho^0$ and $f_0$ obtained from the fit to the
BW$\times$PS function in the hadronic ``cocktail". The $d^2N/(2\pi
p_Tdp_Tdy)$ distributions at mid-rapidity ($|y| \!<\!$ 0.5) as a
function of $p_T$ for minimum bias pp and peripheral Au+Au
collisions are depicted in Fig.~\ref{fig:spectra}. In pp
interactions, a power-law fit was used to extract the $\rho^0$ and
$f_0$ yields per unit of rapidity. In Au+Au collisions, an
exponential fit in $m_T - m_0$ was used to extract the $\rho^0$
and $f_0$ yields and the inverse slopes. Here, $m_0$ is the
$\rho^0$ and $f_0$ masses reported in \cite{15}. The systematic
uncertainty in the $\rho^0$ and $f_0$ yields is estimated to be
15$\%$ and 50$\%$, respectively for both minimum bias pp and
peripheral Au+Au. The systematic errors quoted are due to
uncertainties in the tracking efficiency and the normalization
between the $M_{\pi\pi}$ and the like-sign reference
distributions.

\begin{figure}[htb]
\begin{minipage}[t]{80mm}
\begin{center}
\epsfxsize=2.8in \epsfbox{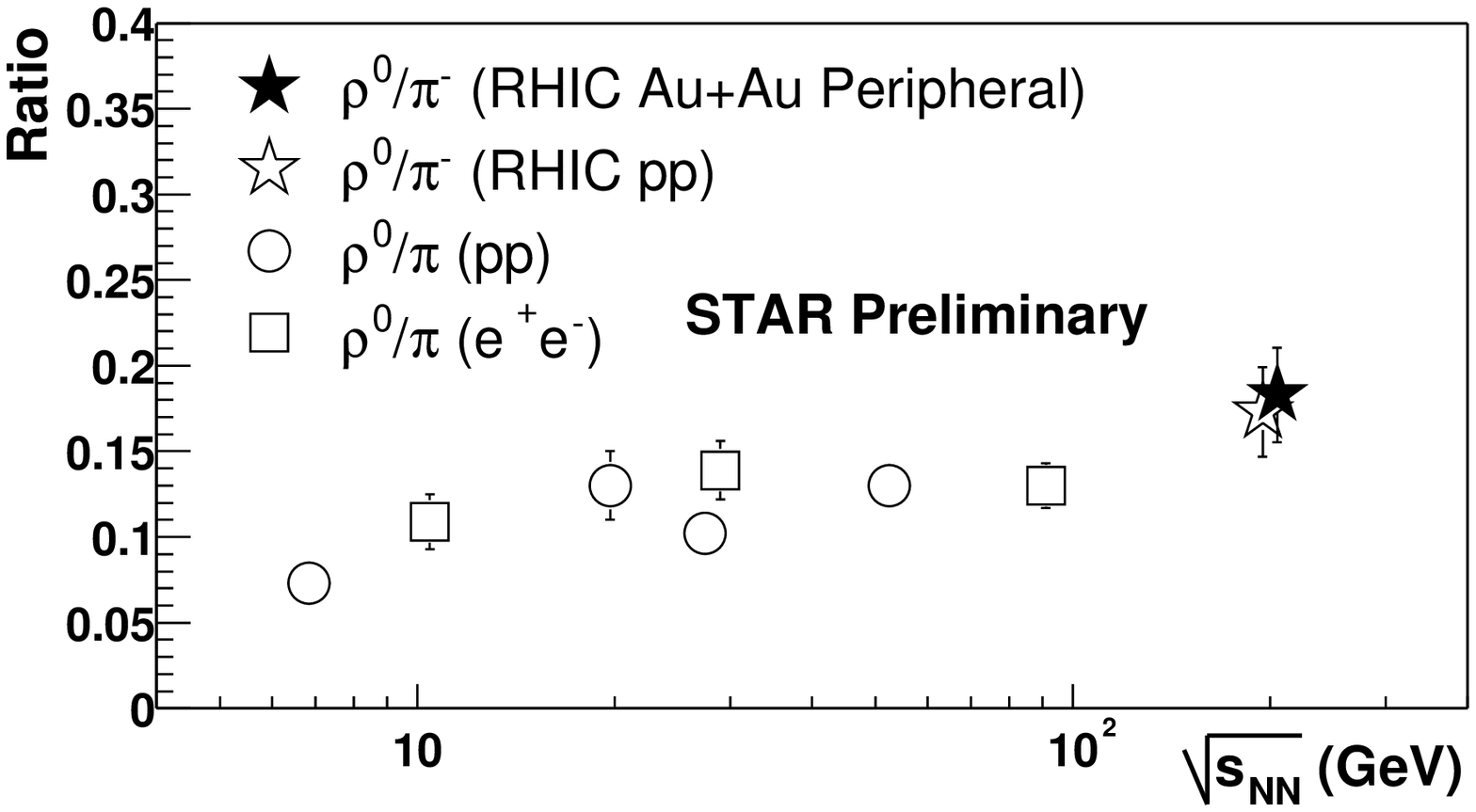}
\end{center}
\end{minipage}
\hspace{\fill}
\begin{minipage}[t]{80mm}
\begin{center}
\epsfxsize=2.85in \epsfbox{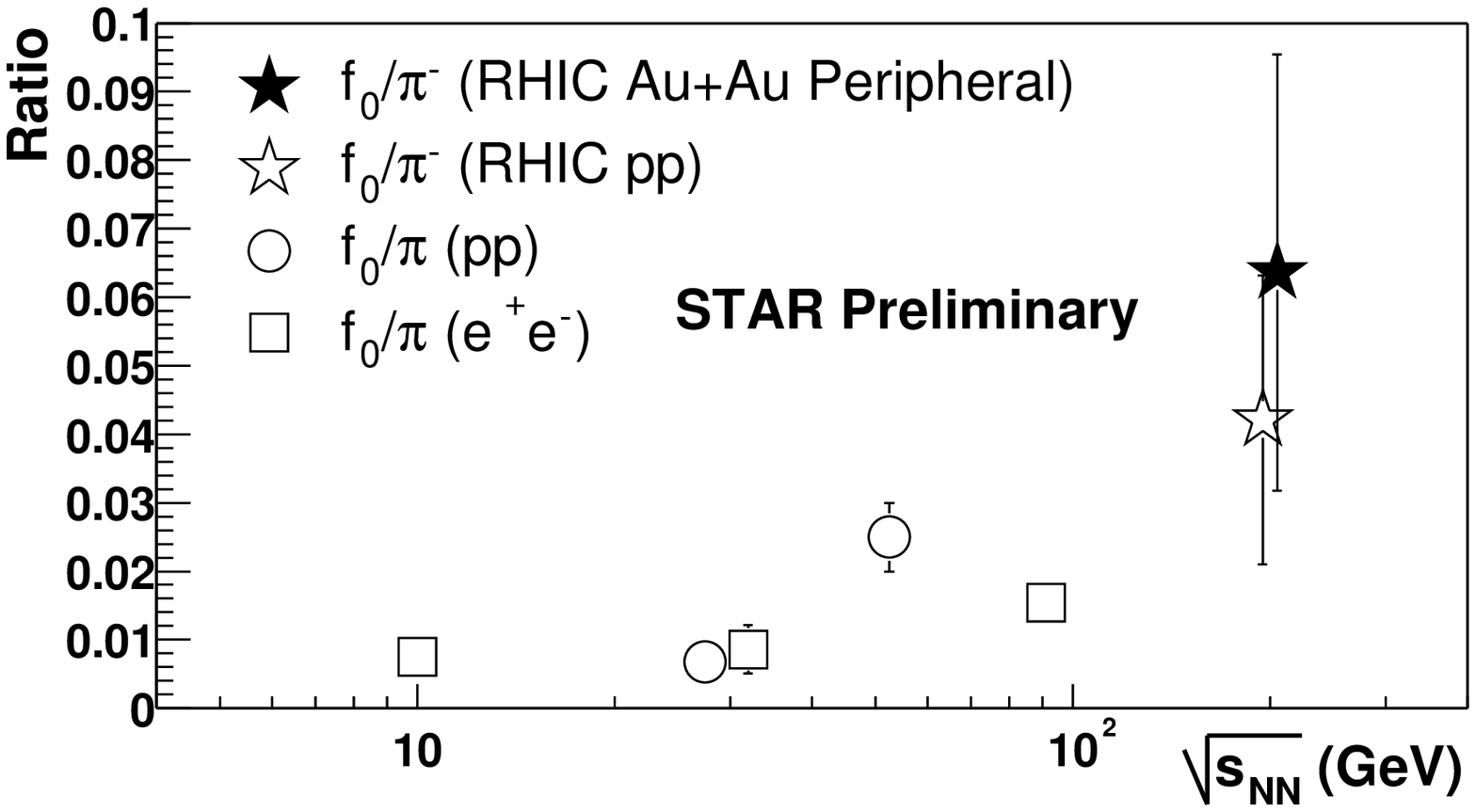}
\end{center}
\end{minipage}
\caption{\label{fig:ratios} $\rho^0/\pi$ (left) and $f_0/\pi$
(right) ratios as a function of beam energy. The ratios from Au+Au
collisions correspond to 40-80$\%$ of the hadronic cross-section.
The ratios are from measurements in e$^+$e$^-$ collisions at 10.45
GeV \cite{22}, 29 GeV \cite{23} and 91 GeV \cite{24} beam energies
and pp at 6.8 GeV \cite{25}, 19.7 GeV \cite{26}, 27.5 GeV
\cite{14}\ and 52.5 GeV \cite{27}. The errors on the ratios at
$\sqrt{s_{NN}}$$\!=\!$ 200 GeV correspond to the quadratic sum of
the statistical and systematic errors, and the $\pi^-$ results are
from Ref. \cite{28}.}
\end{figure}

The $\rho^0/\pi^-$ and $f_0/\pi^-$ ratios as a function of beam
energy for different colliding systems are depicted in
Fig.~\ref{fig:ratios}. The $\rho^0$ production slightly increases
with beam energy and the $\rho^0/\pi^-$ ratios for minimum bias pp
and peripheral Au+Au interactions are interestingly comparable. In
the case of the $f_0$, an analysis with smaller uncertainties is
necessary for a conclusive statement.

\section{Conclusions}
We have presented preliminary results on $\rho(770)^0$ and
$f_{0}(980)$ production at mid-rapidity in minimum bias pp and
peripheral Au+Au collisions at $\sqrt{s_{NN}}$ $\!=\!$ 200 GeV.
This is the first direct measurement of $\rho^0(770)
\!\rightarrow\! \pi^+\pi^-$ and $f_0(980) \!\rightarrow\!
\pi^+\pi^-$ in heavy-ion collisions. The measured $\rho^0$ mass
increases as a function of $p_T$ and is lower than previous
measurements reported in \cite{15} by $\sim$40 MeV/$c^2$ in
minimum bias pp and $\sim$70 MeV/$c^2$ in peripheral Au+Au
collisions. The same behavior was observed by the OPAL experiment
at LEP \cite{11}. Dynamical interactions with the surrounding
matter or the interference between different $\pi^+\pi^-$
scattering channels are possible explanations to the apparent
modification of the $\rho^0$ meson properties. The $\rho^0$
production slightly increases with beam energy and the
$\rho/\pi^-$ ratios in minimum bias pp and peripheral Au+Au are
interestingly comparable. Future measurements of the $\rho^0$
meson in both leptonic and hadronic channels in pA and different
centralities in Au+Au collisions and possibly higher $p_T$
coverage will provide important information on the collision
dynamics of relativistic collisions.

\end{document}